\documentclass[universe,article,accept,moreauthors,pdftex,10pt,a4paper]{mdpi} 
\firstpage{1} 
\pubvolume{5}
\issuenum{1}
\articlenumber{0}
\pubyear{2019}
\copyrightyear{2019}
\history{Received: 31 March 2019; Accepted: 22 May 2019; Published: date}
\newcommand{\ttt}[1]{\texttt{\small #1}}
\updates{yes}
\usepackage{amssymb,amsmath,amsfonts,amssymb,mathrsfs}
\usepackage{lipsum}
\usepackage{hyperref}
\usepackage{bm}  
\usepackage{graphicx}   
\Title{Neutron-Star-Merger Equation of State}

\Author{Veronica Dexheimer $^{1,}$*, Constantinos Constantinou $^{1}$, Elias R. Most  $^{2}$, L. Jens Papenfort $^{2}$, \mbox{Matthias Hanauske $^{2,3}$,}
Stefan Schramm $^{2,3}$, Horst Stoecker $^{2,3,4}$ and Luciano Rezzolla $^{2,3}$}
\AuthorNames{Veronica Dexheimer, Constantinos Constantinou and Firstname Lastname}

\address{%
$^{1}$ \quad Department of Physics, Kent State University, Kent, OH 44242, USA; cconsta5@kent.edu \\
$^{2}$ \quad Institut f{\"u}r Theoretische Physik,
  Max-von-Laue-Stra{\ss}e 1, 60438 Frankfurt, Germany; most@fias.uni-frankfurt.de (E.R.M.);  papenfort@th.physik.uni-frankfurt.de (L.J.P.); hanauske@fias.uni-frankfurt.de (M.H.); schramm@fias.uni-frankfurt.de (S.C.);  h.stoecker@gsi.de (H.S.); rezzolla@itp.uni-frankfurt.de (L.R.)\\
$^{3}$ \quad Frankfurt Institute for Advanced Studies,
  Ruth-Moufang-Stra{\ss}e 1, 60438 Frankfurt, Germany\\
$^{4}$ \quad  GSI Helmholtzzentrum f{\"u}r Schwerionenforschung GmbH, 64291 Darmstadt, Germany}
\corres{Correspondence: vdexheim@kent.edu}

\abstract{In this work, we discuss the dense matter equation of state (EOS) for the extreme range of conditions encountered in neutron stars and their mergers. The calculation of the properties of such an EOS involves modeling different degrees of freedom (such as nuclei, nucleons, hyperons, and quarks), taking into account different symmetries, and including finite density and temperature effects in a thermodynamically consistent manner. We begin by addressing subnuclear matter consisting of nucleons and a small admixture of light nuclei in the context of the excluded volume approach. We then turn our attention to supranuclear homogeneous matter as described by the Chiral Mean Field (CMF) formalism. Finally, we present results from realistic neutron-star-merger simulations performed using the CMF model that predict signatures for deconfinement to quark matter in gravitational wave signals.}


\begin{document}

\setcounter{section}{0} 
\section{Introduction}

The first detection of gravitational waves from the neutron-star-merger GW170817 \cite{Abbott2017_etal}, has generated considerable interest in the equation of state (EOS) of matter created in such extreme events. In these events, temperatures of tens of MeV are expected to be generated over regions that vary in density by several orders of magnitude. Therefore, it is imperative to adopt EOS  descriptions in which the symmetries and degrees of freedom change according to the local temperature and density conditions of the system. From the point of view of merger simulations, it is important to use a small grid for the simulation to capture the complexity of the event. 

In this work, we present a compilation of former works that treated separately the low- and high-density regimes expected to be formed in neutron-star mergers. First, we discuss an extension of the excluded volume (EV) approach, which takes into consideration light nuclei beyond the usual $\alpha$-particle in the subnuclear regime. Here, the EOS of Akmal, Pandharipande, and Ravenhall (APR) is used as the underlying description for interacting nucleons. Second, for the description of the supranuclear regime, we discuss the Chiral Mean Field (CMF) formalism, in which the hadronic degrees of freedom include both nucleons and hyperons.  For sufficiently high temperature and/or density, the CMF model incorporates a first-order phase transition to deconfined quark matter. The merger simulations of this work are performed using a covariant general-relativistic description of hydrodynamics coupled to a fully general-relativistic spacetime evolution. The numerical grid in the simulation achieves the highest resolution of $250$ m covering the two stars and a total \mbox{extent of $1500$~km}.

\section{Subnuclear Density}

In this section, we discuss the subnuclear density region, which we take to be $n \lesssim 0.1$ fm$^{-3}$. Besides density, thermodynamic variables also depend on temperature $T$ and 
electron fraction \mbox{$Y_e=n_e/n$}, the latter being equal to baryonic charge fraction $Y_c=\Sigma_B Q_i n_i/n$ due to the charge neutrality requirement. In this regime, a uniform phase of nucleonic matter would be mechanically (spinodally) unstable, which would give rise to a negative compressibility. Although this instability shrinks in size with increasing isospin asymmetry (lower electron/charge faction), it does not disappear for small temperatures. As a solution to the problem, an inhomogeneous phase must be included in the formalism. In our case, it consists (in addition to nucleons and electrons) of light nuclear clusters. As a result, the mechanical instability is lifted for $T\geq 8$ MeV. At lower temperatures, mechanically stable configurations must necessarily include heavy nuclei ($A>4$) and/or pasta phases.

In the EV approach, the $\alpha$-particles and other light nuclei ($^3$H, and $^3$He, and $^4$He) are assumed to be structureless and their interactions with ``outside'' nucleons are taken into account by treating them as rigid spheres of constant volume. This treatment accounts only for repulsive interactions, the attractive ones being deemed small. The total free energy density can be decomposed as $F = F_b + F_e + F_\gamma$, where the different terms stand for baryon, electron, and photon contributions. The baryon contribution consists of a nuclear contribution given by non-interacting Boltzmann gases and an ``outside'' nucleon contribution given by the APR EoS. 

The APR EOS is a Skyme-like parametric fit \cite{APR98} to the microscopic calculations of Akmal and Pandharipande~\cite{Akmal:1997ft}, where the NN interaction is described by the 
Argonne-18 2-body potential, a modified Urbana-IX 3-body potential (so that the binding energy of isospin-symmetric nuclear matter is $-$16 MeV), and a 2-body relativistic boost potential. The Argonne-18 potential is a high precision fit to the Nijmegen scattering database, such that it reproduces phase shifts and scattering lengths. The Urbana-IX describes 2-pion exchange and includes phenomenological in-medium modifications (involving $\Delta$ isobars) to the 2-body interaction. The boost potential is a correction to the 2-body potential when the interaction is observed in a frame other than the rest-frame of the nucleons. The expectation value of the ground state corresponding to the total potential is determined by variational chain summation techniques using a variational wave-function consisting of a symmetrized product of pair-correlation operators acting on the Fermi gas wave-function. As a result, isospin-symmetric nuclear matter equilibrium bulk properties other than $n_{sat}$ and $E/A(n_{sat})$ are predictions of the model, instead of quantities to which it is fitted.

\begin{figure}[H]
\centering
\includegraphics[width=10cm]{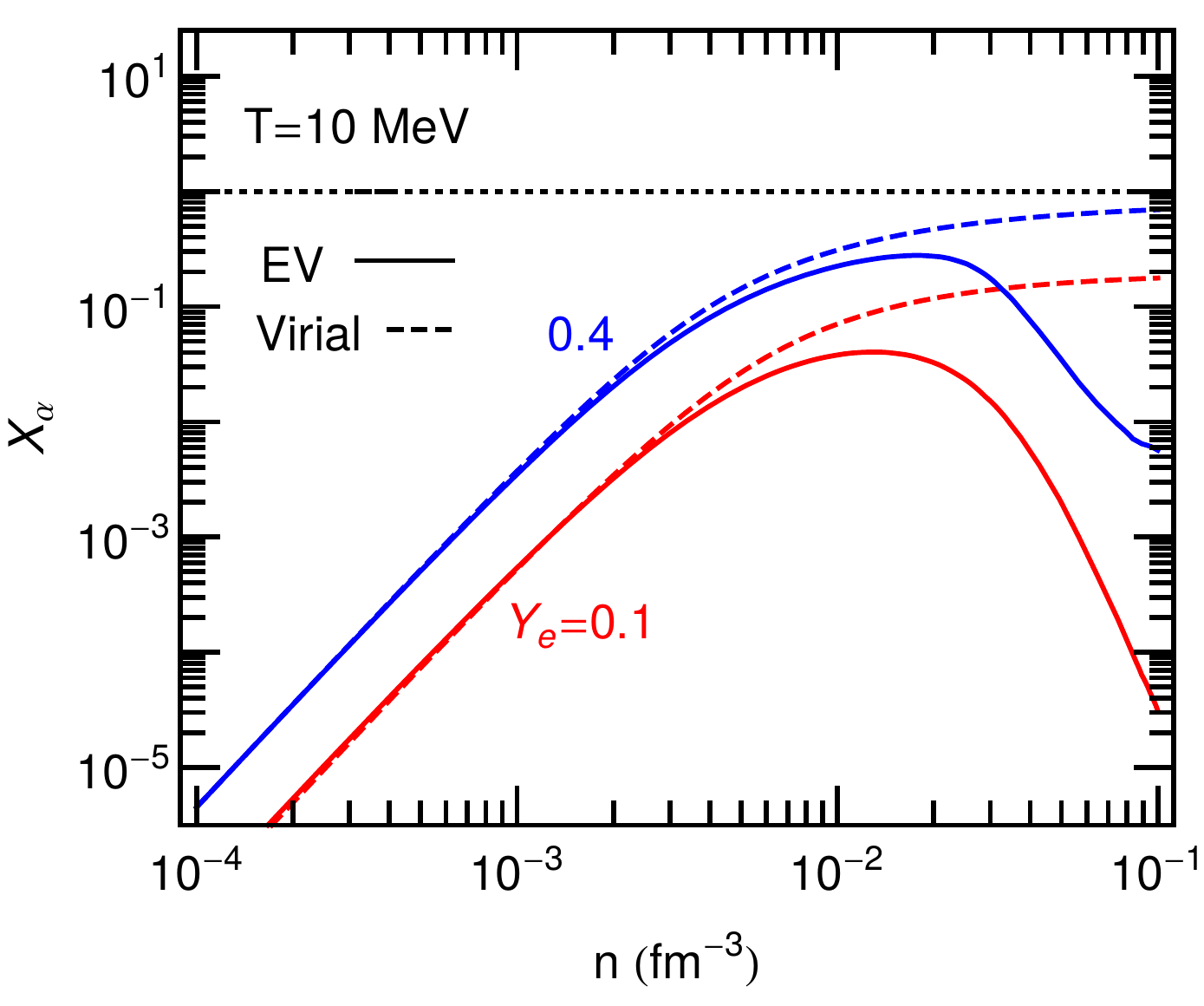}
\caption{$\alpha$-particle mass fraction $X_\alpha=4 n_\alpha/n$ as a function of baryon density for different electron fractions at T = 10 MeV shown for the APR/excluded volume and virial approaches.}
\label{figure1}
\end{figure}

Figure~\ref{figure1}, shows a comparison of EV with virial expansion results for $np\alpha$ matter. The virial approach includes bound and continuum state corrections to the 
ideal gas results for thermal variables~\cite{BU37}. It is model-independent, as experimental phase-shift data are used as input for the theory. The pressure is obtained from the partition function ${\cal Q}$ according to $P=T/V \log{\cal Q}$ and it is expressed in terms of the fugacities $z_i$,  ($i =\alpha$, n, p) and the 2nd virial coefficients $b_2$, which are simple integrals involving thermal weights and elastic scattering phase shifts. As expected, there are fewer $\alpha$ particles at lower lepton fractions (blue vs. red curves). The APR/EV approach predicts significantly lower $\alpha$-particle populations at large densities relative to the virial approach (solid vs dashed curves). This is a consequence of only repulsive 
interactions being included in the EV approach, which is not the case in the virial approach. In the latter, owing to the lack of sufficiently high-energy data such 
that the nuclear hard-core is resolved, the $\alpha$-nucleon interaction is predominantly attractive. The sudden disappearance of the $\alpha$ particles causes a change in slope in the baryonic component of thermodynamic quantities. In the case of lower temperatures, this causes a non-monotonic behavior in the baryonic component of thermodynamic quantities. The non-monotonic behavior is smoothed out when other light nuclei are considered (see Ref.~\cite{Lalit:2018dps}  for more details).

\begin{figure}[H]
\centering
\includegraphics[width=6.5 cm]{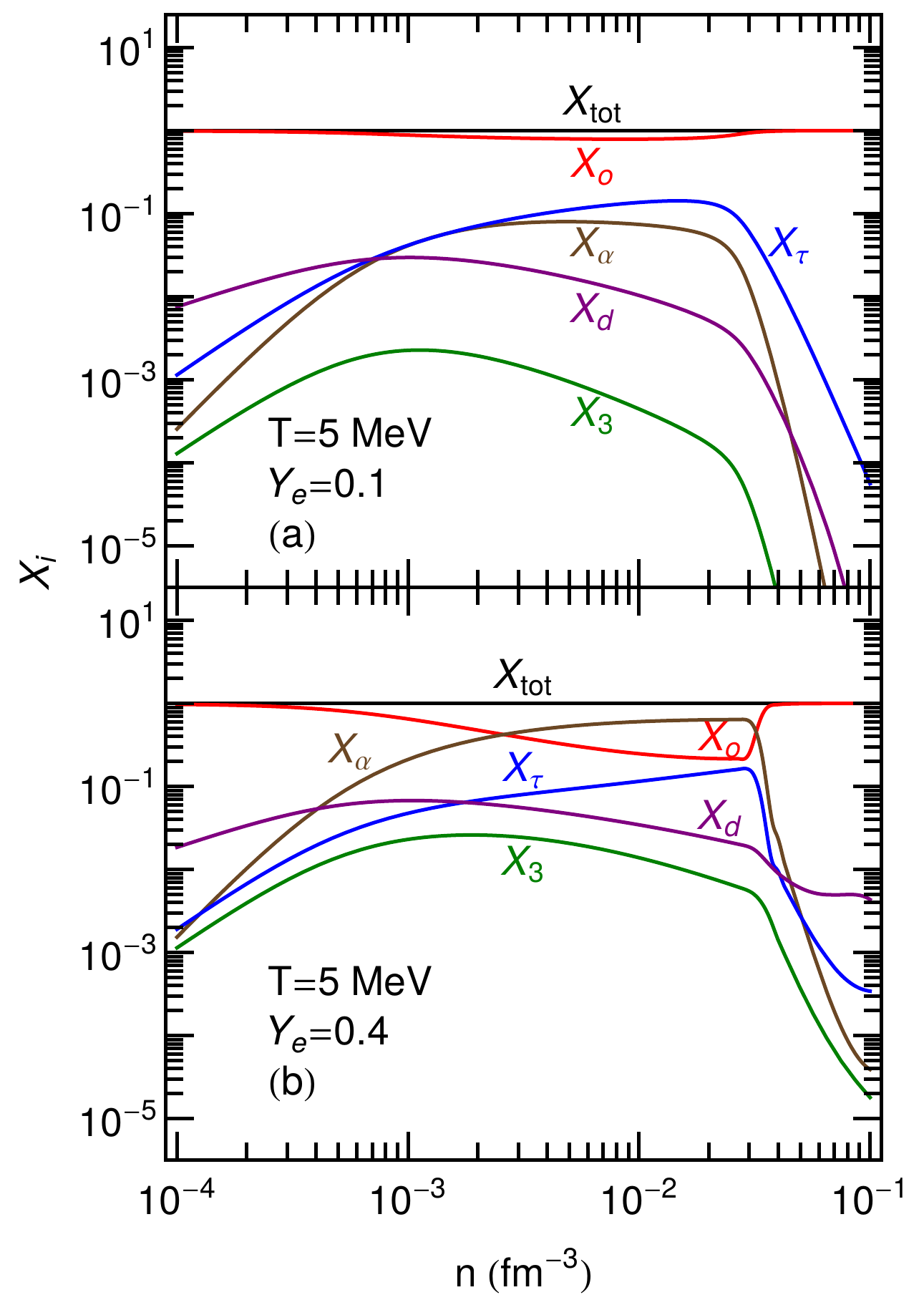}
\includegraphics[width=6.55 cm]{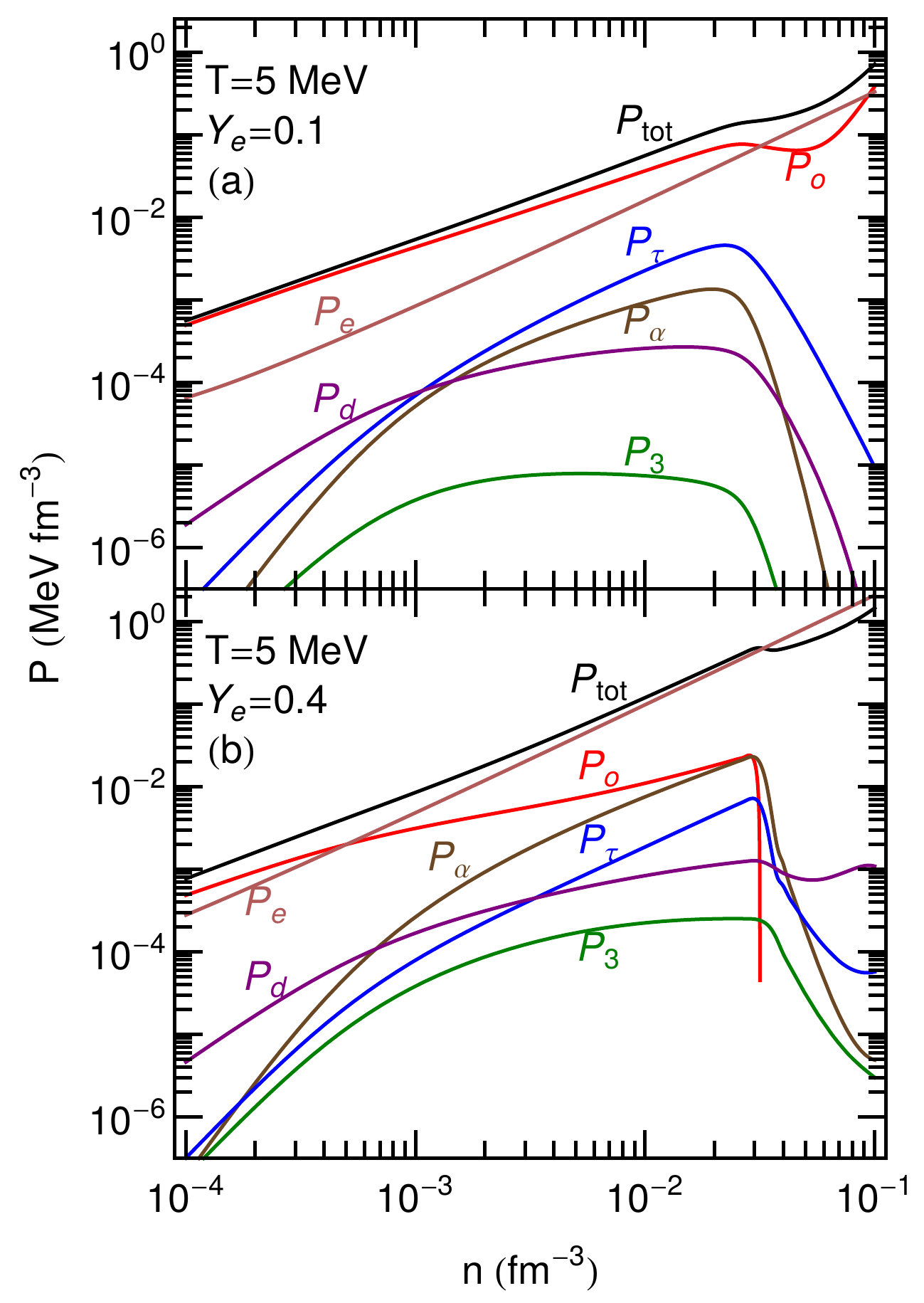}
\caption{Mass fraction $X_i=A_i n_i/n$ (\textbf{left}) and pressure (\textbf{right}) as a function of baryon density for different electron fractions at T = 5 MeV shown for the APR/EV approach. The subscripts $d$ ,$\tau$, 3, $\alpha$, $e$, and $o$ correspond to contributions from deuterons, tritons, $^3$He, alpha particles, electrons and outside nucleons (nucleons not 
bound in nuclei). The total mass fraction (=1) and total pressure are given by $X_{tot}$ and $P_{tot}$, respectively.}
\label {figure2}
\end{figure}

\begin{figure}[H]
\centering
\includegraphics[width=6.5 cm]{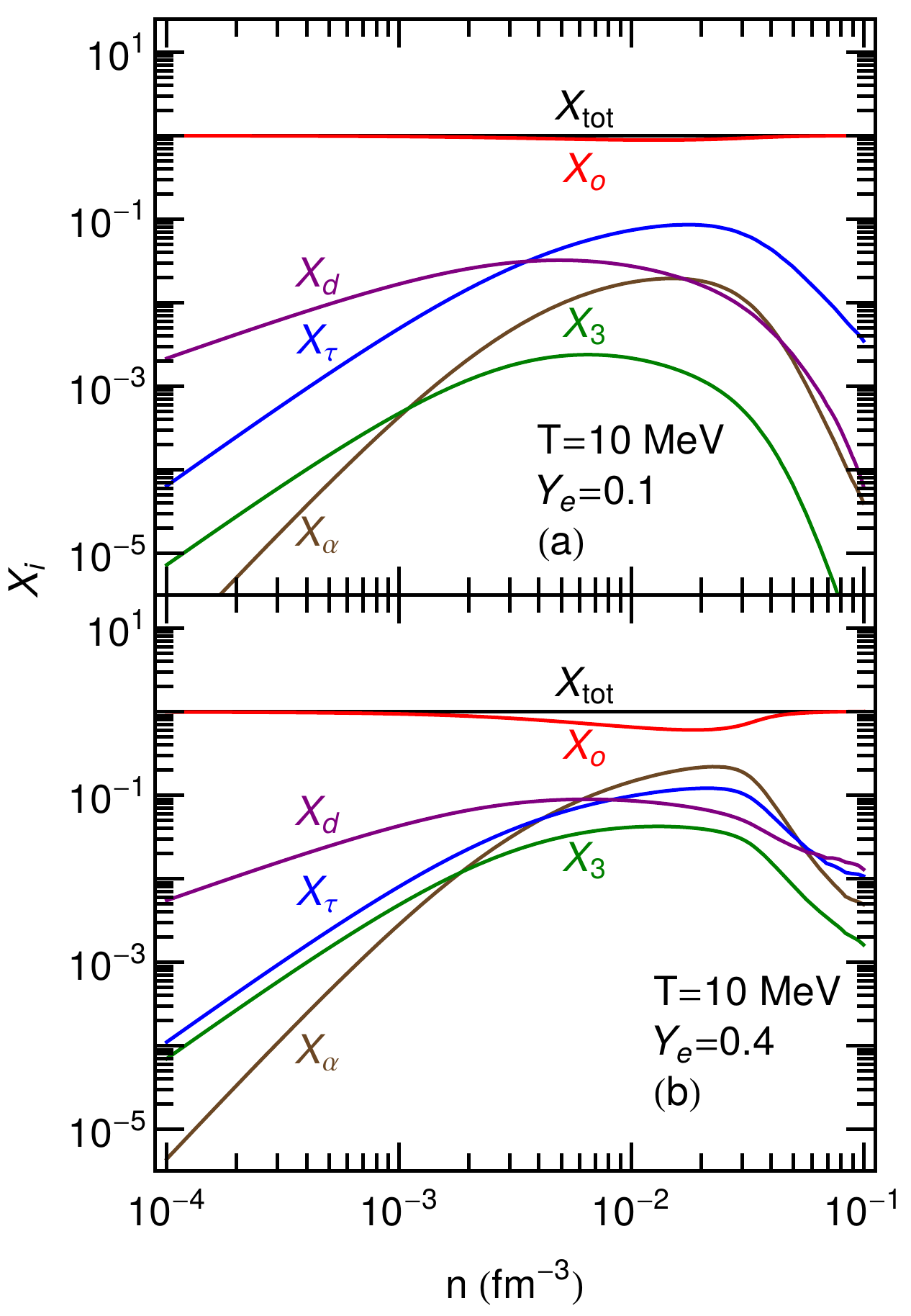}
\includegraphics[width=6.65 cm]{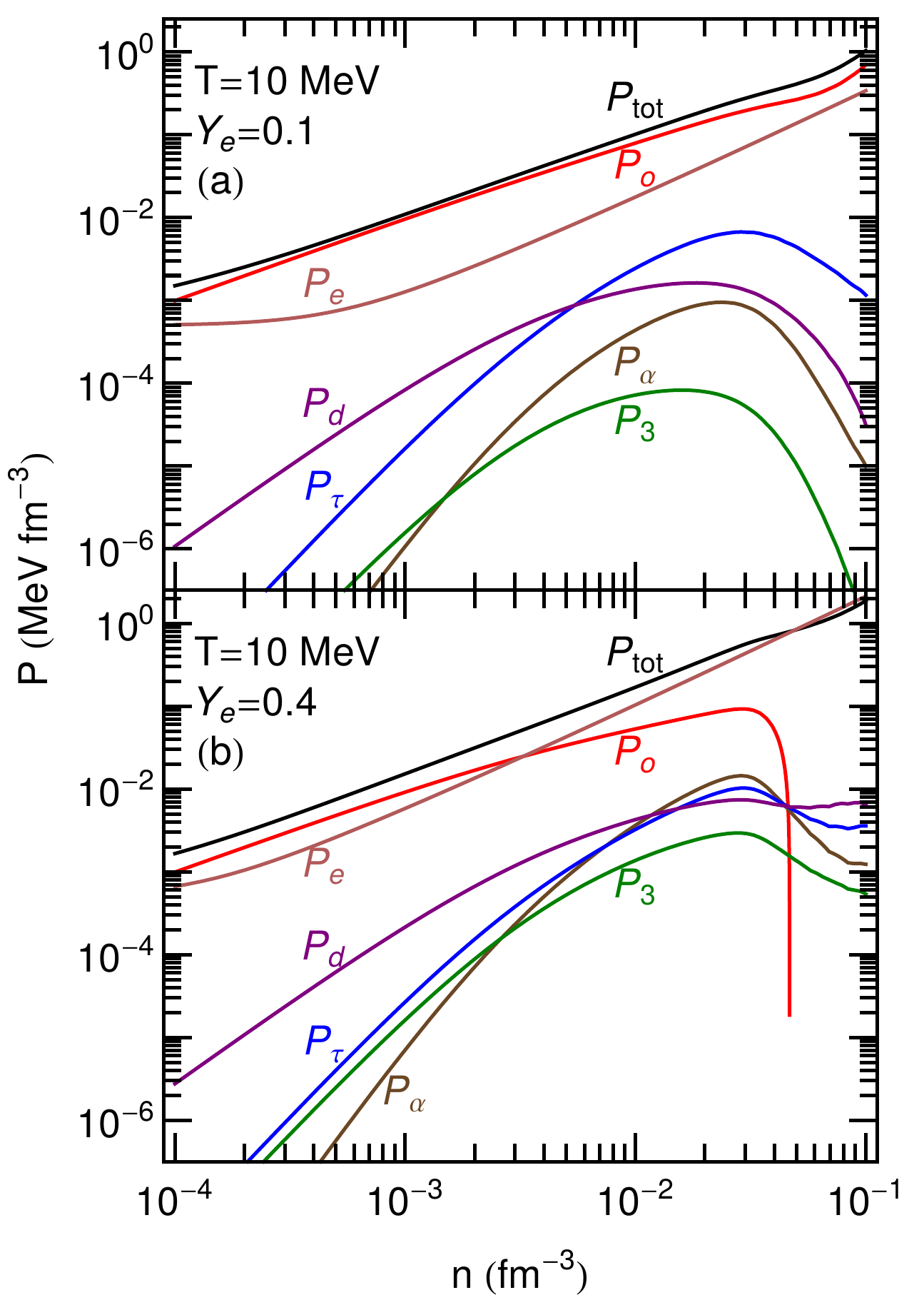}
\caption{Mass fraction $X_i=A_i n_i/n$ (\textbf{left}) and pressure (\textbf{right}) as a function of baryon density for different electron fractions at T=10 MeV shown for the APR/EV approach. The subscripts $d$ ,$\tau$, 3, $\alpha$, $e$, and $o$ correspond to contributions from deuterons, tritons, $^3$He, alpha particles, electrons and outside nucleons (nucleons not 
bound in nuclei). The total mass fraction (=1) and total pressure are given by $X_{tot}$ and $P_{tot}$, respectively.}
\label {figure3}
\end{figure}

The mass fractions of the various light nuclear clusters, as calculated by applying the EV approach to the APR EOS, are presented in the left panels of Figures \ref{figure2} and \ref{figure3} for temperatures of $5$ MeV and $10$ MeV, respectively. These are determined by a combination of the charge and baryon number conservation laws, the binding energy of each type of nucleus, and  local conditions ($n,~Y_e,~T$). The density at which a particular species vanishes is primarily controlled by the EV $v_i$ assigned to it. In this instance, we are using the corresponding experimentally determined charge radii. Note, however, that this choice is by no means compulsory. For example, in Ref. \cite{LLPR85}, $v_{\alpha}$ was obtained by calculating the effective interaction range consistent with an optical potential fitted to neutron-$\alpha$ scattering data. Moreover, the presence of heavy nuclei (not taken into account here) will also affect the relative~particle~concentrations.
\unskip

The right panels of Figures \ref{figure2} and \ref{figure3} show the contributions of light nuclei, nucleons, and electrons to the total pressure. Being that the light nuclei are treated as non-interacting 
gases, their pressures are given by classical ideal gas expressions, $P_i = n_i T$ , modulated by EV factors. The latter act significantly only at the higher end of the density range considered here and cause the populations of the light nuclei to decline. Please note that electronic contributions dominate at higher densities and even more so as $Y_e$ increases (lower panels). The negative slope in the total pressure around 0.03 fm$^{-3}$ in the bottom right panel of Figure~\ref{figure2} is related to the spinodal instability of purely nucleonic matter (and, therefore, the nuclear liquid-gas phase transition) which is more pronounced at lower temperatures and for more isospin-symmetric configurations. This unphysical behavior is not present in the bottom right panel of Figure~\ref{figure3}. Even though the APR liquid-gas critical temperature of about 18~MeV has not been reached in this case, mainly due to electrons, and to a lesser extend to the light nuclei, the  total pressure remains monotonically increasing at $10$ MeV temperature.
\unskip
\unskip

\section{Supranuclear Density}

For densities larger than $n \sim 0.1~{\rm fm}^{-3}$, matter is too dense for nuclei of any type to form and, thus, consists of uniformly distributed nucleons and electrons. On 
one hand, effective field theory has already enabled first-principle calculations of isospin-symmetric and asymmetric matter with systematic corrections. On the other, 
continuing beyond $n \sim 0.3~{\rm fm}^{-3}$ to encompass the central densities of neutron stars is precluded in these methods, as the perturbative expansion parameter reaches 
uncomfortably large values (see Ref.~\cite{Tews:2019cap} for a recent review of the topic). Phenomenological approaches based on non-relativistic potential models with contact and finite-range interactions have long been used to explore the EOS at 
supranuclear densities, the advantage of these models being that calculations are relatively easier than first-principle calculations. However, higher-than-two-body interactions, found necessary to fit constraints offered by laboratory data on nuclei at near-nuclear densities, can render these EOS's acausal at large density due to the lack of Lorentz invariance in a non-relativistic approach. These higher-than-two-body interactions correspond to terms in the energy density which vary as $n^{\sigma}$ with $\sigma >2$. At 
sufficiently high densities, they will dominate all other contributions including the thermal parts and lead to superluminal behavior $(c_s/c)^2\simeq \sigma-1 >1$.

As a solution to this problem, relativistic Dirac-Brueckner-Hartree-Fock \cite{MPA87,Engvik96,Baldo_2012,Baldo2001} and mean field-theoretical~\cite{MS96} models are used at supranuclear 
densities, as they and their extensions are inherently Lorentz invariant and, thus, preserve causality. We choose to work with the CMF model, which is based on a nonlinear realization of the SU(3) sigma model \cite{Papazoglou:1998}. This framework incorporates chiral symmetry and its restoration at large densities and temperatures, as predicted in QCD. Being that hadrons and quarks in the CMF model interact via meson exchange in a chirally invariant manner, the various particle masses originate from interactions with the medium. The model in this specific parametrization is in agreement with standard nuclear and astrophysical constraints \cite{Dexheimer:2008, Negreiros:2010}, as well as lattice QCD and perturbative QCD \cite{Dexheimer:2009, Roark:2018}. In particular, in the limit of zero-temperature and zero-angular momentum, it predicts a maximum mass of $2.1\,M_{\odot}$ for a hadronic star and $2.0\,M_{\odot}$ when quarks are included.

\begin{figure}[H]
\centering
\includegraphics[width=7.2 cm]{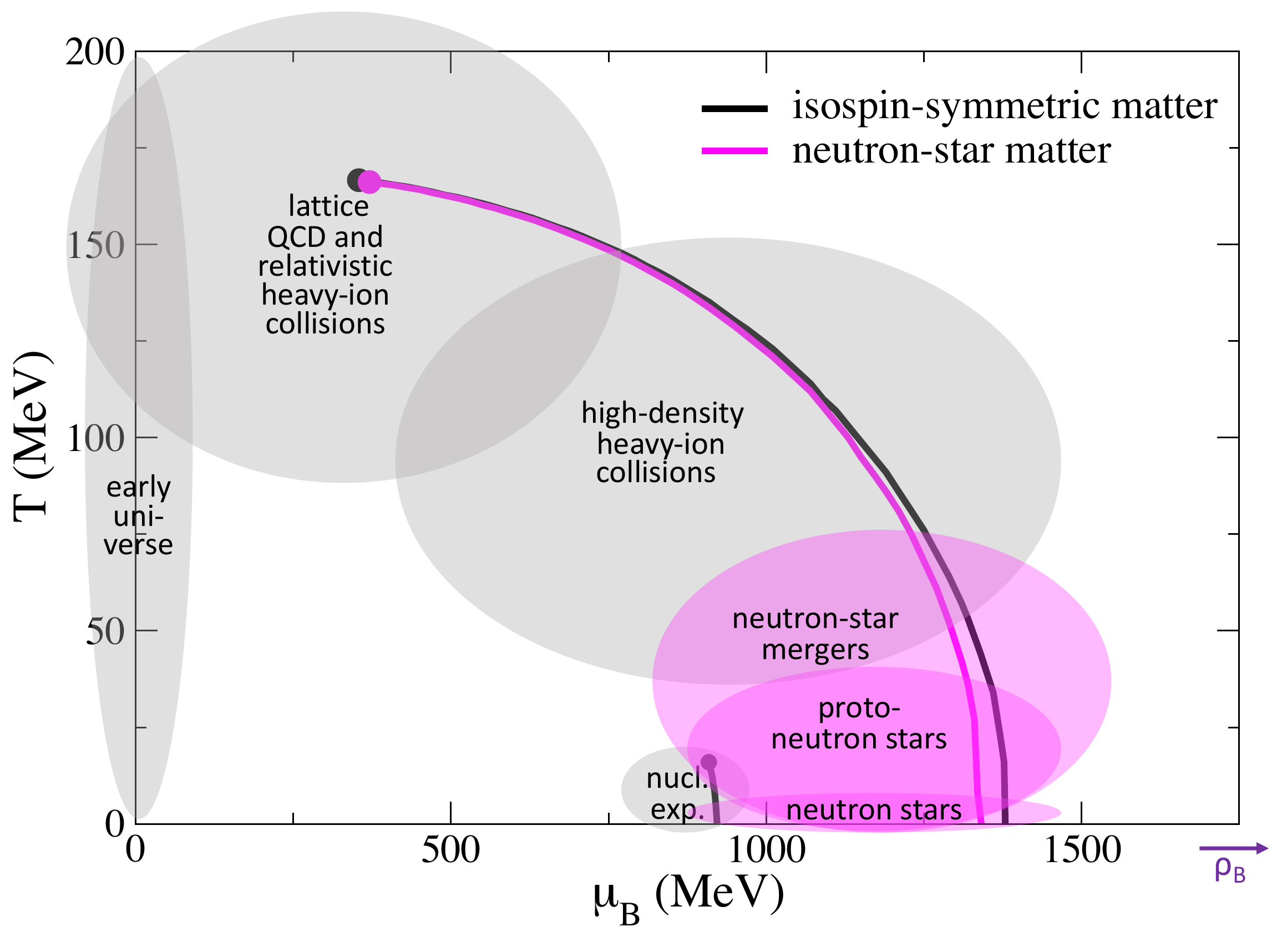}
\includegraphics[trim= 0 0.7cm 0 0,clip,width=8 cm]{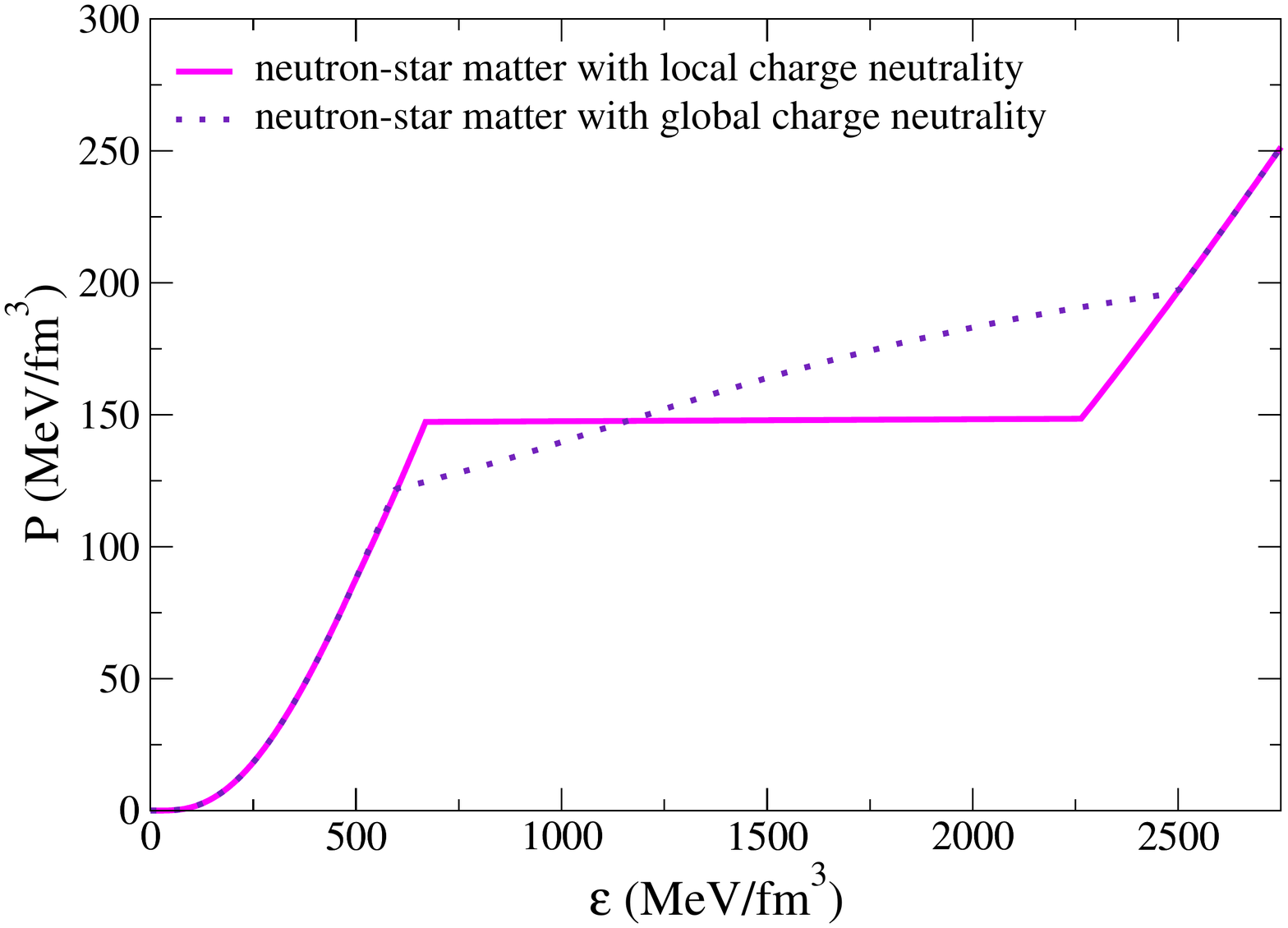}
\caption{\textbf{Left}: QCD Phase diagram resulting from the CMF model. The lines represent first-order transitions. The circles mark the critical end-points. Isospin-symmetric matter 
refers to zero isospin and strangeness constraints, while neutron-star matter stands for charged neutral matter in chemical equilibrium. The shaded regions exemplify some of the 
different regimes that can be described within the model. \textbf{Right}: EoS for star matter at T = 0 under different charge neutrality conditions calculated with the CMF model.}
\label{figure4}
\end{figure}

This approach allows for the existence of soluted quarks in the hadronic phase and soluted hadrons in the quark phase at finite temperature. However, quarks (hadrons) will always 
give the dominant contribution in the quark (hadron) phase, and the two phases can be distinguished by their approximate order parameters, e.g., the chiral condensate $\sigma$ for chiral symmetry restoration or the field $\Phi$ for deconfinement (named in analogy with the Polyakov loop). This inter-penetration of quarks and hadrons (that 
increases with temperature) provides a physically effective description and is indeed required to achieve the crossover transition known to take place at small chemical potential
values \cite{Aoki:2006}. The left panel of Figure~\ref{figure4} contains the QCD phase diagram (modified from Ref.~\cite{Dexheimer:2017ecc}) resulting from the CMF model and illustrates these features. The right panel in Figure~\ref{figure4} shows the 
neutron-star matter EOS (assuming charge neutrality and chemical equilibrium) at zero temperature calculated from the CMF model. It can be seen that if the local charge neutrality condition is relaxed to a global charge neutrality condition, a mixed phase appears. In the following, we {\bf are not} going to allow for such relaxation (equivalent to considering a large surface tension between the phases) to study the maximum effect the phase transition can have in binary mergers. We are also going to use a neutrino-leakage scheme to evolve the charge fraction of matter, which will allow us to go beyond the initially chemically equilibrated data.
\unskip

The neutron-star-merger simulations \cite{Most:2018eaw} discussed next are performed using the  \ttt{Frankfurt/IllinoisGRMHD} code (\ttt{FIL})  \citep{DelZanna2007,Borges2008,Harten83,Galeazzi2013} including weak-interactions via the neutrino-leakage scheme ~\cite{Ruffert96b, Rosswog:2003b, OConnor10}. The binaries are initially placed at a distance of $45$ km in quasi-circular orbit and perform around five orbits before the merger. These simulations include two setups with equal-mass neutron stars with a combined total mass of $M=2.8$ and $2.9\, M_\odot$. For each of these systems, two identical scenarios were simulated either employing the standard CMF EOS, where quarks and a strong first-order PT are included, or a purely 
hadronic variant, in which the quarks are artificially~suppressed. 

\begin{figure}[H]
\centering
\includegraphics[width=7 cm]{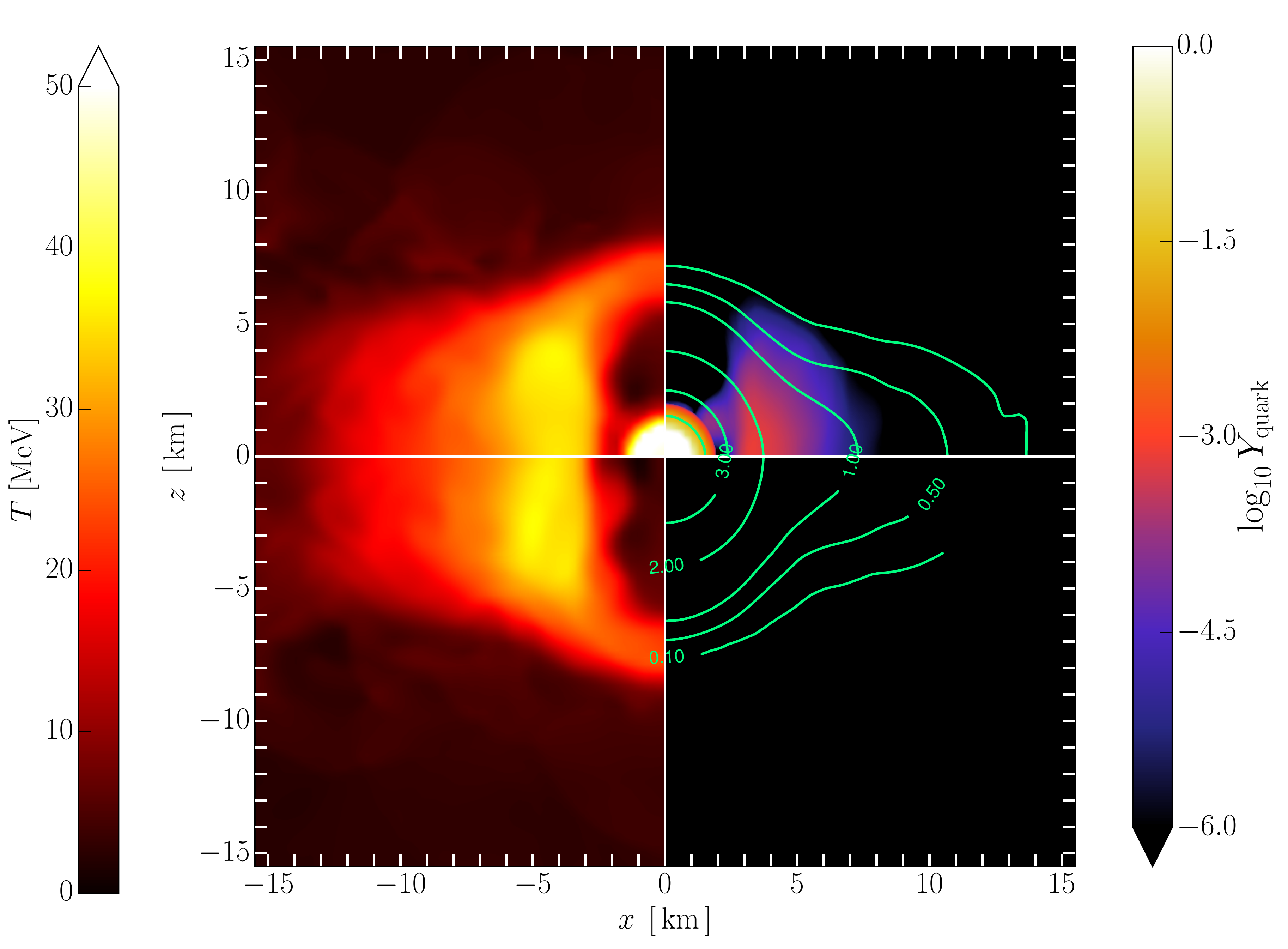}
\includegraphics[width=8.5 cm]{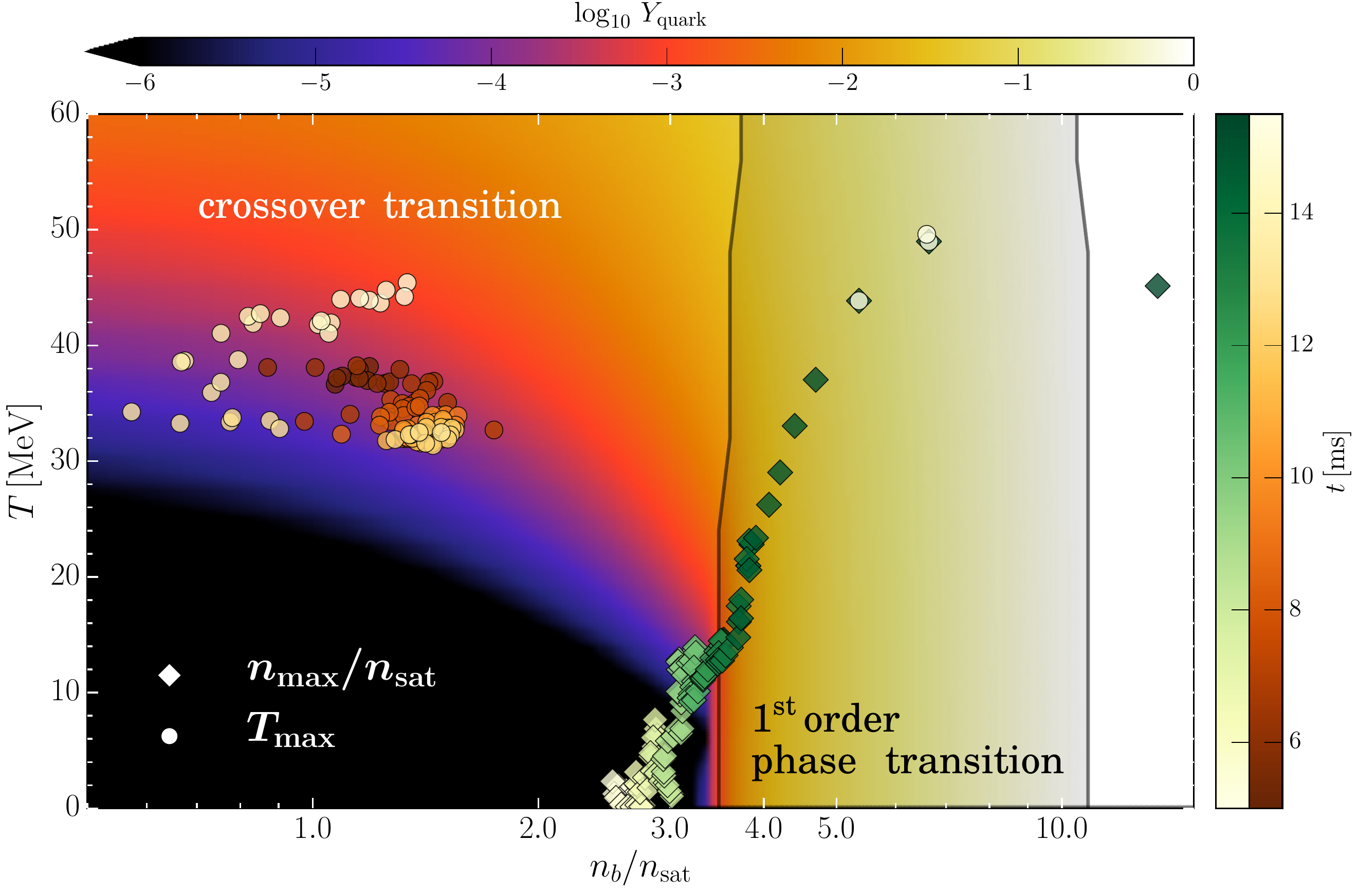}
\caption{{\bf Left}: merger simulations performed using the CMF model without (top) and with the suppression of quarks (bottom) for a high-mass binary at a time shortly before the 
collapse to a black hole (figure extracted from Ref.~\cite{Most:2018eaw}). {\bf Right}: time evolution of the maximum normalized baryon density region (diamonds) and maximum temperature region (circles) after the merger for the 
low-mass binary using the CMF EoS. The gray-shaded area shows the first-order PT being crossed.}
\label{figure5}
\end{figure}

The left panel of Figure \ref{figure5} shows the meridional plane for the $2.9$ M$_\odot$ binary $7.7$ ms after the merger, when the first-order phase transition has already occurred and formed a hot and dense core inside the hypermassive neutron star. Different 
subpanels compare simulations performed with the CMF model allowing for quarks (top subpanels) or artificially suppressing quarks (bottom subpanels). The top subpanels show
that a large quark fraction is only present in the center and outside ring, where the temperature is high. Please note that in the bottom subpanels, due to the lack of a first-order PT having taken place, there is no hot central region. This feature is a consequence of the sudden compactification generated by the very steep first-order phase transition and would have been significantly less pronounced
if a mixture of phases had been included in the EOS. 

The right panel of Figure \ref{figure5} shows which parts of the EOS and the QCD phase diagram are actually probed between $5$ ms and $15$  ms after the merger for the low-mass binary 
remnant.  The diamonds show the evolution of the maximum baryon density, which basically probes the center of the merged object. The circles show the evolution of the maximum 
temperature, which probes different regions of the remnant but, eventually, coincides with the center (when circles and diamonds meet). The continued emission of GWs and, hence, the induced loss of angular momentum through GWs 
leads to a continuous rise of the central density, which ultimately reaches and crosses the boundary of the first-order PT (gray-shaded area).

In the right panel of Figure \ref{figure5}, the background color code illustrates the relative fraction of quarks compared with the total baryonic one throughout the whole phase diagram. Although it is not shown, the first-order phase transition region will bend left at large temperatures (as indicated in the 
left panel of Figure~\ref{figure4}) but will eventually disappear reaching the crossover region. The first-order phase transition appears in the right panel of Figure~\ref{figure5} as a surface because we show the phase diagram as a function of density and the reason there are points inside the region is (i) due to the way our data was slightly interpolated in that range and (2) the small but finite resolution of the dynamical time-scale of the hydrodynamical simulation. The first-order phase transition would correspond to a line if we had chosen to plot the phase diagram as a function of chemical potential instead. This and other plots will be discussed in a separate publication. Please note that as discussed in detail in Ref.~\cite{Most:2018eaw}, the deconfinement phase transition is enough to produce a dephasing in the gravitational wave form after the merger. 
This is different from hyperon effects or early quark deconfinement, which should leave imprints in the waveform already before or immediately after the merger 
\cite{Sekiguchi2011b, Radice2017a,Bauswein:2018bma}. 

\section{Discussion and Outlook}

Despite advances made in dense nuclear matter theory, many issues remain unresolved and/or poorly constrained with attendant uncertainties in simulations of supernovae and 
binary mergers of neutron stars. In this work, we explored the effects of light nuclear clusters such as d, $^3$H, $^3$He and $^4$He on the properties of the EOS at subnuclear
densities. Their interactions among themselves and with nucleons were described in the EV approach. We found that the light cluster relative concentrations are mostly dependent upon the 
binding energy of the various species. The volume associated with each type of nucleus plays a role only at the higher density range of the subnuclear region and characterizes 
the population decline of the species at those densities. 

We also emphasized that while the EV approach is very useful, it accounts only for repulsive interactions among nuclei and nucleons. We know that attractive 
interactions are present from phase-shift data and their inclusion in a calculation will certainly provide corrections to the system's state variables. Nevertheless, we expect 
that these corrections are likely to be small below densities of about 0.01 fm$^{-3}$, as indicated from a comparison with the virial approach.

Considering supranuclear densities, the CMF model represents a very useful tool to study the influence of different hyperonic and quark degrees 
of freedom in astrophysical scenarios. This can be seen, for example, in signatures of deconfinement phase transition predicted to exist in neutron-star mergers.  Although 
the differences generated by quarks in this case are not large, they will be possible to resolve if detected by third-generation GW detectors 
\cite{Punturo:2010, Evans2016}, especially in the case that a merger takes place in a nearby system.

While the CMF approach is suitable to describe matter at supranuclear densities (in the neutron-star core), another description is needed for subnuclear 
densities (the crust and the very low-density regions produced in binary mergers). For the moment, we have matched the CMF EOS to the nuclear statistical equilibrium description 
presented in Ref.~\cite{daSilvaSchneider2017}, but we are soon going to connect it to the EV approach discussed in this proceeding \cite{Lalit:2018dps}. This complete 
table will be made publicly available at CompOSE.

\vspace{6pt}
\authorcontributions{All authors contributed significantly to this work.} 




\funding{Support comes
from ``PHAROS'' COST Action CA16214, LOEWE-Program in HIC for
FAIR, European Union's Horizon 2020 Research and Innovation Programme
(Grant 671698) (call FETHPC-1-2014, project ExaHyPE), ERC Synergy
Grant ``BlackHoleCam: Imaging the Event Horizon of Black Holes'' (Grant
No. 610058), Judah M. Eisenberg - Laureatus Professur at the fachbereich physik at Goethe Universitaet, and the National Science Foundation under grant
PHY-1748621. The simulations were performed on the SuperMUC cluster at
the LRZ in Garching, on the LOEWE cluster in CSC in Frankfurt, and on the
HazelHen cluster at the HLRS~in~Stuttgart.}

\conflictsofinterest{The authors declare no conflict of interest.}


\reftitle{References}


\begin{thebibliography}{-------}
\providecommand{\natexlab}[1]{#1}

\end{thebibliography}


\begin{thebibliography}{999}
\providecommand{\natexlab}[1]{#1}


\bibitem[Abbott \em{et~al.}(2017)Abbott et~al.]{Abbott2017_etal}
Abbott, B.P.; Abbott, R.; Abbott, T.D.; Acernese, F.; Ackley,~K.; Adams, C.; Adams, T.; Addesso, P.; Adhikari,~R.X.; Adya, V.B.; et al.
\newblock GW170817: Observation of Gravitational Waves from a Binary Neutron
  Star Inspiral.
\newblock {\em Phys. Rev. Lett.} {\bf 2017}, {\em 119},~161101.

\bibitem[Akmal \em{et~al.}(1998)Akmal, Pandharipande, and Ravenhall]{APR98}
Akmal, A.; Pandharipande, V.R.; Ravenhall, D.G. Equation of state of nucleon matter and neutron star structure.
\newblock {\em Phys. Rev. C} {\bf 1998}, {\em 58},~1804.

\bibitem[Akmal and Pandharipande(1997)]{Akmal:1997ft}
Akmal, A.; Pandharipande, V.R.
\newblock {Spin - isospin structure and pion condensation in nucleon matter}.
\newblock {\em \mbox{Phys. Rev. C}} {\bf 1997}, {\em 56},~2261--2279.

\bibitem[Beth and Uhlenbeck(1937)]{BU37}
Beth, E.; Uhlenbeck, G.E. The quantum theory of the non-ideal gas. II. Behaviour at low temperatures. 
\newblock {\em Physica} {\bf 1937}, {\em 4},~915--924.

\bibitem[Lalit \em{et~al.}(2019)Lalit, Mamun, Constantinou, and
  Prakash]{Lalit:2018dps}
Lalit, S.; Mamun, M.A.A.; Constantinou, C.; Prakash, M.
\newblock {Dense matter equation of state for neutron star mergers}.
\newblock {\em Eur. Phys. J. A} {\bf 2019}, {\em 55},~10.


\bibitem[Lamb \em{et~al.}(1985)Lamb, Lattimer, Pethick, and Ravenhall]{LLPR85}
Lamb, D.Q.; Lattimer, J.M.; Pethick, C.J.; Ravenhall, D.G. Physical properties of hot, dense matter: The general case.
\newblock {\em Nucl. Phys. A} {\bf 1985}, {\em 432},~646--742.

\bibitem[Tews \em{et~al.}(2019)Tews, Margueron, and Reddy]{Tews:2019cap}
Tews, I.; Margueron, J.; Reddy, S.
\newblock {Confronting gravitational-wave observations with modern nuclear
  physics constraints}. \emph{arXiv} \textbf{2019}, arXiv:1901.09874.

\bibitem[M\"{u}ther \em{et~al.}(1987)M\"{u}ther, Prakash, and Ainsworth]{MPA87}
M\"{u}ther, H.; Prakash, M.; Ainsworth, T.L. 
	The nuclear symmetry energy in relativistic Brueckner-Hartree-Fock calculations.
\newblock {\em Phys. Lett.} {\bf 1987}, {\em B199},~469--474.

\bibitem[Engvik \em{et~al.}(1996)Engvik, Hjorth-Jensen, Osnes, Bao, and
  Ostgaard]{Engvik96}
Engvik, L.; Hjorth-Jensen, M.; Osnes, E.; Bao, G.; Ostgaard, E. Asymmetric Nuclear Matter and Neutron Star Properties. 
\newblock {\em Astrophys. Jl.} {\bf 1996}, {\em 469},~794.

\bibitem[Baldo and Burgio(2012)]{Baldo_2012}
Baldo, M.; Burgio, G.F.
\newblock Properties of the nuclear medium.
\newblock {\em Rep. Prog. Phys.} {\bf 2012}, {\em 75},~026301.

\bibitem[Baldo and Burgio(2001)]{Baldo2001}
Baldo, M.; Burgio, F., Microscopic Theory of the Nuclear Equation of State and
  Neutron Star Structure.
\newblock In {\em Physics of Neutron Star Interiors}; Blaschke, D., Sedrakian,
  A., Glendenning, N.K., Eds.; Springer: Berlin/Heidelberg, German, 2001; pp. 1--29.

\bibitem[M\"{u}ller and Serot(1996)]{MS96}
M\"{u}ller, H.; Serot, B.D. Relativistic mean-field theory and the high-density nuclear equation of state.
\newblock {\em \mbox{Nucl. Phys. A}} {\bf 1996}, {\em 606},~508--537.

\bibitem[Papazoglou \em{et~al.}(1999)Papazoglou, Zschiesche, Schramm,
  Schaffner-Bielich, Stoecker, and Greiner]{Papazoglou:1998}
Papazoglou, P.; Zschiesche, D.; Schramm, S.; Schaffner-Bielich, J.; Stoecker,
  H.; Greiner, W.
\newblock {Nuclei in a chiral SU(3) model}.
\newblock {\em Phys. Rev.  C} {\bf 1999}, {\em 59},~411--427.

\bibitem[Dexheimer and Schramm(2008)]{Dexheimer:2008}
Dexheimer, V.; Schramm, S.
\newblock {Proto-Neutron and Neutron Stars in a Chiral SU(3) Model}.
\newblock {\em Astrophys. J.} {\bf 2008}, {\em 683},~943--948.

\bibitem[Negreiros \em{et~al.}(2010)Negreiros, Dexheimer, and
  Schramm]{Negreiros:2010}
Negreiros, R.; Dexheimer, V.A.; Schramm, S.
\newblock {Modeling Hybrid Stars with an SU(3) non-linear sigma model}.
\newblock {\em Phys. Rev. C} {\bf 2010}, {\em 82},~035803.

\bibitem[Dexheimer and Schramm(2010)]{Dexheimer:2009}
Dexheimer, V.A.; Schramm, S.
\newblock {A Novel Approach to Model Hybrid Stars}.
\newblock {\em Phys. Rev. C} {\bf 2010}, {\em 81},~045201.

\bibitem[Roark and Dexheimer(2018)]{Roark:2018}
Roark, J.; Dexheimer, V.
\newblock {The Deconfinement Phase Transition in Proto-Neutron-Star Matter.}
 \emph{Phys. Rev. C}  \textbf{2018}, \emph{26}, 055805.

\bibitem[Aoki \em{et~al.}(2006)Aoki, Endrodi, Fodor, Katz, and
  Szabo]{Aoki:2006}
Aoki, Y.; Endrodi, G.; Fodor, Z.; Katz, S.D.; Szabo, K.K.
\newblock {The Order of the quantum chromodynamics transition predicted by the
  standard model of particle physics}.
\newblock {\em Nature} {\bf 2006}, {\em 443},~675--678.

\bibitem[Dexheimer \em{et~al.}(2017)Dexheimer, Hempel, Iosilevskiy, and
  Schramm]{Dexheimer:2017ecc}
Dexheimer, V.; Hempel, M.; Iosilevskiy, I.; Schramm, S.
\newblock {Phase transitions in dense matter}.
\newblock {\em Nucl. Phys. A} {\bf 2017}, {\em 967},~780--783.

\bibitem[Most \em{et~al.}(2019)Most, Papenfort, Dexheimer, Hanauske, Schramm,
  Stocker, and Rezzolla]{Most:2018eaw}
Most, E.R.; Papenfort, L.J.; Dexheimer, V.; Hanauske, M.; Schramm, S.; Stocker,
  H.; Rezzolla, L.
\newblock {Signatures of quark-hadron phase transitions in general-relativistic
  neutron-star mergers}.
\newblock {\em Phys. Rev. Lett.} {\bf 2019}, {\em 122},~061101.

\bibitem[{Del Zanna} \em{et~al.}(2007){Del Zanna}, {Zanotti}, {Bucciantini},
  and {Londrillo}]{DelZanna2007}
{Del Zanna}, L.; {Zanotti}, O.; {Bucciantini}, N.; {Londrillo}, P.
\newblock {ECHO: A Eulerian conservative high-order scheme for general
  relativistic magnetohydrodynamics and magnetodynamics}.
\newblock {\em Astron. Astrophys.} {\bf 2007}, {\em 473},~11--30.

\bibitem[Borges \em{et~al.}(2008)Borges, Carmona, Costa, and Don]{Borges2008}
Borges, R.; Carmona, M.; Costa, B.; Don, W.
\newblock {An improved weighted essentially non-oscillatory scheme for
  hyperbolic conservation laws}.
\newblock {\em J. Comput. Phys.} {\bf 2008}, {\em
  227},~3191--3211.

\bibitem[Harten \em{et~al.}(1983)Harten, Lax, and van Leer]{Harten83}
Harten, A.; Lax, P.D.; van Leer, B.
\newblock On Upstream Differencing and Godunov-Type Schemes for Hyperbolic
  Conservation Laws.
\newblock {\em SIAM Rev.} {\bf 1983}, {\em 25},~35.

\bibitem[{Galeazzi} \em{et~al.}(2013){Galeazzi}, {Kastaun}, {Rezzolla}, and
  {Font}]{Galeazzi2013}
{Galeazzi}, F.; {Kastaun}, W.; {Rezzolla}, L.; {Font}, J.A.
\newblock {Implementation of a simplified approach to radiative transfer in
  general relativity}.
\newblock {\em Phys. Rev. D} {\bf 2013}, {\em 88},~064009.

\bibitem[{Ruffert} \em{et~al.}(1996){Ruffert}, {Janka}, and
  {Schaefer}]{Ruffert96b}
{Ruffert}, M.; {Janka}, H.T.; {Schaefer}, G.
\newblock {Coalescing neutron stars---A step towards physical models. I.
  Hydrodynamic evolution and gravitational-wave emission.}
\newblock {\em Astron. Astrophys.} {\bf 1996}, {\em 311},~532--566.

\bibitem[{Rosswog} and {Liebend{\"o}rfer}(2003)]{Rosswog:2003b}
{Rosswog}, S.; {Liebend{\"o}rfer}, M.
\newblock {High-resolution calculations of merging neutron stars-II. Neutrino
  emission}.
\newblock {\em Mon. Not. R. Astron. Soc.} {\bf 2003}, {\em 342},~673--689.

\bibitem[{O'Connor} and {Ott}(2010)]{OConnor10}
{O'Connor}, E.; {Ott}, C.D.
\newblock {A new open-source code for spherically symmetric stellar collapse to
  neutron stars and black holes}.
\newblock {\em Class. Quantum Grav.} {\bf 2010}, {\em 27},~114103.

\bibitem[{Sekiguchi} \em{et~al.}(2011){Sekiguchi}, {Kiuchi}, {Kyutoku}, and
  {Shibata}]{Sekiguchi2011b}
{Sekiguchi}, Y.; {Kiuchi}, K.; {Kyutoku}, K.; {Shibata}, M.
\newblock {Effects of Hyperons in Binary Neutron Star Mergers}.
\newblock {\em \mbox{Phys. Rev. Lett.}} {\bf 2011}, {\em 107},~211101.

\bibitem[{Radice} \em{et~al.}(2017){Radice}, {Bernuzzi}, {Del Pozzo},
  {Roberts}, and {Ott}]{Radice2017a}
{Radice}, D.; {Bernuzzi}, S.; {Del Pozzo}, W.; {Roberts}, L.F.; {Ott}, C.D.
\newblock {Probing Extreme-density Matter with Gravitational-wave Observations
  of Binary Neutron Star Merger Remnants}.
\newblock {\em Astrophys. J. Lett.} {\bf 2017}, {\em 842},~L10.

\bibitem[Bauswein \em{et~al.}(2019)Bauswein, Bastian, Blaschke, Chatziioannou,
  Clark, Fischer, and Oertel]{Bauswein:2018bma}
Bauswein, A.; Bastian, N.U.F.; Blaschke, D.B.; Chatziioannou, K.; Clark, J.A.;
  Fischer, T.; Oertel, M.
\newblock {Identifying a first-order phase transition in neutron star mergers
  through gravitational waves}.
\newblock {\em Phys. Rev. Lett.} {\bf 2019}, {\em 122},~061102.

\bibitem[Punturo \em{et~al.}(2010)Punturo et~al.]{Punturo:2010}
Punturo, M.; Abernathy, M.; Acernese, F.; Allen, B.; Andersson, N.; Arun, K.; Barone, F.; Barr, B.; Barsuglia,~M.; Beker, M.; et al. 
\newblock {The third generation of gravitational wave observatories and their
  science reach}.
\newblock {\em \mbox{Class. Quant. Grav.}} {\bf 2010}, {\em 27},~084007.

\bibitem[Abbott \em{et~al.}(2017)Abbott et~al.]{Evans2016}
Abbott, B.P.; Abbott, R.; Abbott, T.D.; Abernathy, M.R.; Ackley, K.; Adams, C.; Addesso, P.; Adhikari, R.X.; Adya, V.B.,; Affeldt, C.; et al.
\newblock {Exploring the Sensitivity of Next Generation Gravitational Wave
  Detectors}.
\newblock {\em \mbox{Class. Quant. Grav.}} {\bf 2017}, {\em 34},~044001.

\bibitem[{da Silva Schneider} \em{et~al.}(2017){da Silva Schneider}, {Roberts},
  and {Ott}]{daSilvaSchneider2017}
{Da Silva Schneider}, A.; {Roberts}, L.F.; {Ott}, C.D.
\newblock {A New Open-Source Nuclear Equation of State Framework based on the
  Liquid-Drop Model with Skyrme Interaction}.
\newblock {\em arXiv} {\bf 2017}, arXiv:1707.01527v2.

\end{thebibliography}
\end{document}